# Benchmarking Contemporary Deep Learning Hardware and Frameworks: A Survey of Qualitative Metrics

Wei Dai
*Department of Computer Science*
*Southeast Missouri State University*
Cape Girardeau, MO, USA
wdai@semo.edu

Daniel Berleant
*Department of Information Science*
*University of Arkansas at Little Rock*
Little Rock, AR, USA
jdberleant@ualr.edu

*Abstract*—This paper surveys benchmarking principles, machine learning devices including GPUs, FPGAs, and ASICs, and deep learning software frameworks. It also reviews these technologies with respect to benchmarking from the perspectives of a 6-metric approach to frameworks and an 11-metric approach to hardware platforms. Because MLPerf is a benchmark organization working with industry and academia, and offering deep learning benchmarks that evaluate training and inference on deep learning hardware devices, the survey also mentions MLPerf benchmark results, benchmark metrics, datasets, deep learning frameworks and algorithms. We summarize seven benchmarking principles, differential characteristics of mainstream AI devices, and qualitative comparison of deep learning hardware and frameworks.

## I. INTRODUCTION

After developing for about 75 years, deep learning technologies are still maturing. In July 2018, Gartner, an IT research and consultancy company, pointed out that deep learning technologies are in the Peak-of-Inflated-Expectations (PoIE) stage on the Gartner Hype Cycle diagram [1] as shown in Figure 2, which means deep learning networks trigger many industry projects as well as research topics [2][3][4].

Image data impacts accuracy of deep learning algorithms. The accuracy of deep learning algorithms can be improved by feeding high quality images into these algorithms. Well-known image sets useful for this include CIFAR-10 [5], MNIST[6], ImageNet [7], and Pascal Visual Object Classes (P-VOC) [8]. The CIFAR-10 dataset has 10 groups, and all images are 32×32 color images. MNIST has digital handwriting images, and these images are black and white. ImageNet and P-VOC are high quality image datasets, and are broadly used in visual object category recognition and detection.

Benchmarking is useful in both industry and academia. The definition from the Oxford English Dictionary [9] states that a benchmark is "To evaluate or check (something) by comparison with an established standard." Deep neural learning networks are leading technologies that extend their computing performance and capability based on flexibility, distributed architectures, creative algorithms, and large volume datasets. Comparing them via benchmarking is increasingly important.

Even though previous research papers provide knowledge of deep learning, it is hard to find a survey discussing qualitative benchmarks for machine learning hardware devices and deep learning software frameworks as shown in Figure 1. In this paper we introduce 11 qualitative benchmarking metrics for hardware devices and six metrics for software frameworks in deep learning, respectively. The paper also provides qualitative benchmark results for major deep learning devices, and compares 18 deep learning frameworks.

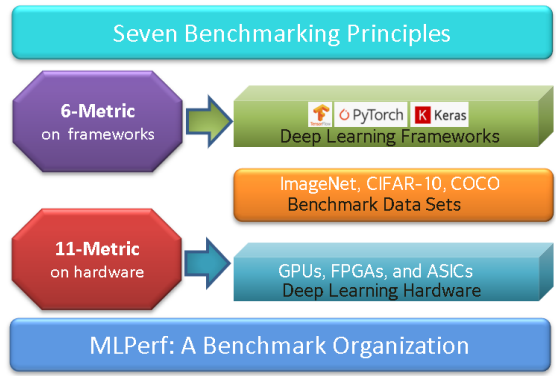

Fig. 1. Benchmarking Metrics and AI Architectures

According to [16],[17], and [18], there are seven vital characteristics for benchmarks. These key properties are:

1) *Relevance*: Benchmarks should measure important features.
2) *Representativeness*: Benchmark performance metrics should be broadly accepted by industry and academia.
3) *Equity*: All systems should be fairly compared.
4) *Repeatability*: Benchmark results should be verifiable.
5) *Cost-effectiveness*: Benchmark tests should be economical.
6) *Scalability*: Benchmark tests should measure from single server to multiple servers.
7) *Transparency*: Benchmark metrics should be readily understandable.

Evaluating artificial intelligence (AI) hardware and deep learning frameworks allows discovering both strengths and weaknesses of deep learning technologies. To illuminate, this paper is organized as follows. Section II reviews hardware platforms including CPUs, GPUs, FPGAs, and ASICs, qualitative metrics for benchmarking hardware, and qualitative results on benchmarking devices. Section III introduces qualitative metrics for benchmarking frameworks and results. Section IV introduces a machine learning benchmark organization named MLPerf and their deep learning benchmarking metrics. Section V presents our conclusions. Section VI discusses future work.

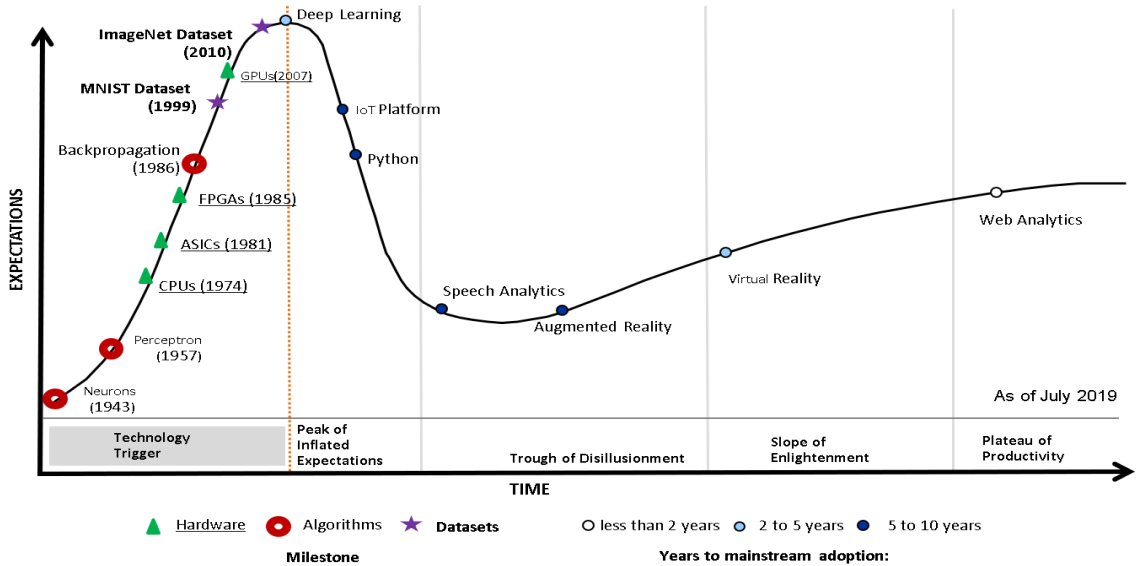

Fig. 2 Milestones of Deep learning on the Gartner hyper cycle. We inserted some deep learning historical milestones, modifying the figure of Gartner [1].

## II. Deep Learning Hardware

AI algorithms often benefit from many-core hardware and high bandwidth memory, in comparison to many non-AI algorithms that are often encountered. Thus computational power is not just a one-dimensional concept. The type of computations the hardware design is best suited for must be considered, since a hardware platform can have more or less computational power depending on the type of computation on which it is measured. GPUs (graphics processing units) do well on the kind of parallelism often beneficial to AI algorithms, in comparison to CPUs (central processing units), and thus tend to be well suited to AI applications. FPGAs (field programmable gate arrays), being configurable, can be configured to perform well on AI algorithms although currently they lack the rich software layer needed to fully achieve their potential in the AI domain. ASICs (application specific integrated circuits) are similar to FPGAs in this regard, since in principle a specially configured FPGA is a kind of ASIC. Thus GPUs, FPGAs and ASICs have the potential to expedite machine learning algorithms in part because of their capabilities for parallel computing and high-speed internal memory.

Nevertheless, while earlier generation CPUs have had performance bottlenecks while training or using deep learning algorithms, cutting edge CPUs can provide better performance and thus better support for deep learning algorithms. In 2017, Intel released Intel Xeon Scalable processors, which includes Intel Advance Vector Extension 512 (Intel AVX-512) instruction set and Intel Math Kernel Library for Deep Neural Networks (Intel MKL-DNN) [10]. The Intel AVX-512 and MKL-DNN accelerate deep learning algorithms on lower precision tasks. Comparing mainstream 32-bit floating point precision (fp32) on GPUs, the 16-bit and 8-bits floating-point precisions (fp16/fp8) are lower in precision, but can be sufficient for the inference of deep learning application domain. In addition, Lower precision also can enhance usage of cache and memory, and can maximize memory bandwidth. Let us look specifically at GPUs, FPGAs, and ASICs next.

### A. GPU Devices

GPUs are specified unitary processors that are dedicated to accelerating real time three-dimensional (3D) graphics. GPUs contain an internal cache, high speed bandwidth, and quick parallel performance. The GPU cache accelerates matrix multiplication routines because these routines do not need to access global memory.

GPUs are universal hardware devices for deep learning. After testing neural networks including ones with 200 hidden layers on MNIST handwritten data sets, GPU performance was found to be better than CPUs [11]. The test results show NVIDIA GeForce 6800 Ultra has a 3.3X speedup compared to the Intel 3GHz P4; ATI Radeon X800 has 2.4-3.4X speedup. NVIDIA GPUs increase FLOPS (**f**loating **p**oint **o**perations **p**er **s**econd) performance. In [12], a single NVIDIA GeForce 8800 GTX, released in November 2006, had 575 CUDA cores with 345.6 gigaflops, and its memory bandwidth was 86.4 GB/s; by September 2018, a NVIDIA GeForce RTX 2080 Ti [13] had 4,352 CUDA cores with 13.4 Teraflops, and its memory bandwidth had increased to 616 GB/s.

### B. FPGA Devices

FPGAs have dynamical hardware configurations, so hardware engineers developed FPGAs using hardware description language (HDL), including VHDL or Verilog [14][15]. However, some use cases will always involve energy-sensitive scenarios. FPGA devices offer better performance per watt than GPUs. According to [16], while comparing gigaflops per watt, FPGA devices often have a 3-4X speed-up compared to GPUs. After comparing performances of FPGAs and GPUs [17] on ImageNet 1K data sets, Ovtcharov et al. [18] confirmed that the Arria 10 GX1150 FPGA devices handled about 233 images/sec. while device power is 25 watts. In comparison, NVIDIA K40 GPUs handled 500-824 images/sec. with device power of 235 watts. Briefly, [17] demonstrates FPGAs can process 9.3 images/joule, but these GPUs can only process 2.1-3.4 images/joule.

## C. ASIC Devices

Usually, ASIC devices have high throughout and low energy consumption because ASICs are fabricated chips designed for special applications instead of generic tasks. While testing AlexNet, one of the convolutional neural networks, the Eyeriss consumed 278 mW [18]. Furthermore, the Eyeriss achieved 125.9 images/joule (with batch size N=4) [19]. In [12], Google researchers confirm that the TPU 1.0, based on ASIC technologies, has about 15-30X speed-up compared to GPUs or CPUs during the same period, with TOPS/watt of about 30-80X better.

## D. Enhance Hardware Performance

Even though multiple cores, CPUs, and hyper-threading are mainstream technologies, these technologies still show weaknesses in the big data era. For example, deep learning models usually have products and matrix transpositions [11], so that these algorithms require intensive computing resources. GPUs, FPGAs, and ASICs have better computing performance with lower latency than conventional CPUs because these specialized chipsets consist of many cores and on-chip memory. The memory hierarchy on these hardware devices is usually separated into two layers: 1) off-chip memory, named global memory or main memory; and 2) on-chip memory, termed local memory or shared memory. After copying data from global memory, deep learning algorithms can use high-speed shared memory to expedite computing performance. Specific program libraries provide dedicated application programming interfaces (APIs) for hardware devices, abstract complex parallel programming, and increased executive performance. For instance, the CuDNN library, released by NVIDIA, can improve performance of the Apache MXNet and the Caffe on NVIDIA GPUs [20][17].

Traditionally, multiple cores, improved I/O bandwidth, and increased core clock speed can improve hardware speeds [21]. In Figure 3, Arithmetic Logic Unit (ALU), single instruction, multiple data (SIMD), and single instruction, multiple thread (SIMT) systems concurrently execute multiply-accumulate (MACs) tasks based on shared memory and configuration files.

However, there are new algorithms to improve computing performance. GPUs are low-latency temporary storage architectures, so the Toeplitz matrix, fast Fourier transform (FFT), and Winograd and Strassen algorithms can be used for improving GPU performance [21]. Data movement consumes energy. FPGAs and ASICs are spatial architectures. These devices contain low-energy on-chip memory, so that reusable dataflow algorithms provide solutions for reducing data movements. Weight stationary dataflow, output stationary dataflow, no local reuse dataflow, and row stationary dataflow were developed for decreasing energy consumption of FPGAs and ASICs [21]. In addition, co-design of deep learning algorithms and hardware devices are other approaches. According to [21], there are two solutions. 1) Decrease precision: There are several algorithms to decrease precision of operations and operands of DNN, such as 8-bit fixed point, binary weight sharing, and log domain quantization. 2) Reduce number of operations and model size: Some algorithms need to be highlighted, such as exploiting activation statistics, network pruning algorithms, and knowledge distillation algorithms.

## E. Qualitative Benchmarking Metrics on Machine Learning Hardware

GPUs, FPGAs, and ASICs can be used in different domains besides deep learning, including cloud servers and edge devices. There are 11 qualitative benchmarking metrics we distinguish on machine learning devices, as follows. In addition, results of the benchmarks are shown in Table I.

TABLE I. QUALITATIVE BENCHMARKING HARDWARE FOR MACHINE LEARNING ([10]-[20])

| # | Attributes | ASICs | FPGAs | GPUs |
|---|---|---|---|---|
| 1 | Computing Performance | High | Low | Moderate |
| 2 | Low Latency | High | Moderate | Low |
| 3 | Energy efficiency | High | Moderate | Good |
| 4 | Compatibility | Low | Moderate | High |
| 5 | Research Costs | High | Moderate | Low |
| 6 | Research Risks | High | Low | Moderate |
| 7 | Upgradability | Low | Moderate | High |
| 8 | Scalability | High | Low | Moderate |
| 9 | Chip Price | Low | High | Moderate |
| 10 | Ubicomp | Low | High | High |
| 11 | Time-to-Market | Low | High | High |

1) *Computing Performance can be measured by FLOPS. For measuring ASICs and GPUs, a quadrillion (thousand trillion) FLOPS (petaflops) are used in testing modern chipsets. In May 2017, Google announced Tensor Processor Unit 2.0 (TPU 2.0), which provides 11.5 petaflops per chip [22]. TPU 3.0, released in May 2018, offers 23.0 petaflops [23]. However, NVIDIA GeForce RTX 2080 Ti has only 13.4 Teraflops [13]. According to [24] and [25], ASICs have the most FLOPs, and GPUs are better than FPGAs.*
2) *Low latency describes an important chipset capability [26], and is distinguished from throughout [12]. In [12][24], ASICs have the lowest latency, while FPGAs are lower than GPUs.*
3) *Energy efficiency in computing is particularly important for edge nodes because mobile devices generally have limited power. In [12][24] ASICs have the highest energy efficiency, and FPGAs and GPUs come in second and third, respectively.*
4) *Compatibility means devices can be supported by multiple deep learning frameworks and popular programming languages. FPGAs needs specially developing libraries, so that FPGAs are not that good with respect to compatibility. GPUs have the best compatibilities [24]. ASICs currently are*

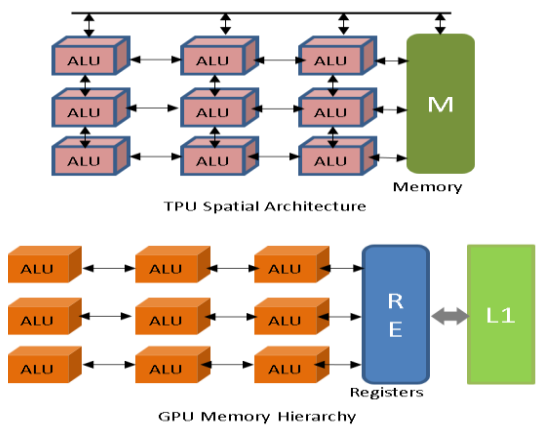

Fig. 3. Parallel Chipsets and memory diagrams (after [21])

second. For example, TPUs support TensorFlow, cafe, etc.

5) Research costs refer to the total costs for developing devices incurred from designing architectures, developing algorithms, and deploying chip sets on hardware devices. GPUs are affordable devices [24]. ASICs are expensive, and FPGAs are between GPUs and ASICs.
6) Research risks are determined by hardware architectures, development risks, and deployed chip sets. ASICs have the highest risks before market scaling. FPGAs are very flexible, so that their risks are limited. GPUs are in the middle.
7) Upgradability is a challenge for most hardware devices. In [24], GPUs are the most flexible after deployment, and are better than FPGAs. ASICs are the most difficult to update after delivery.
8) Scalability means hardware devices can scale up quickly with low costs. Scalability is vital for clouds and data centres. ASICs have excellent scalability. GPUs have good scalability, but not as good as ASICs. FPGAs are the lowest on this dimension.
9) Chip Price means price of each unit chip after industrial-scale production. In [27], FPGAs have the highest chip cost after production scale-up. ASICs have the lowest cost, and GPUs are in the middle.
10) Ubicomp (also named ubiquitous computing) indicates hardware devices used extensively for varied use cases including e.g. large scale clouds and low energy mobile devices. FPGAs are very flexible, so that the devices can be used in different industries and scientific fields. ASICs usually are dedicated to specific industry needs. GPUs like FPGAs can be developed for many research fields and industry domains.
11) Time-to-market means the length of time from design to sale of products. According to [15],[24], and [27], FPGAs and GPUs have lower development time than ASICs.

III. MAINSTREAM DEEP LEARNING FRAMEWORKS

Open source deep learning frameworks allow engineers and scientists to define activation functions, develop special algorithms, train big data, and deploy neural networks on different hardware platforms, from x86 servers to mobile devices.

Based on the wide variety of usages, support teams, and development interfaces, we split 18 frameworks into three sets including mature frameworks, developing frameworks, and inactive frameworks. The 10 mature frameworks can be used currently to enhance training speed, improve scalable performance, and reduce development risks. The developing frameworks are not yet broadly used in industries or research projects, but some developing frameworks could be used in specific fields. Retired frameworks are largely inactive.

A. Mature Frameworks

1) Caffe and Facebook Caffe2: Caffe [28] was developed at the University of California, Berkeley in C++. According to [29], Caffe can be used on FPGA platforms. Caffe 2 [30] is an updated framework supported by Facebook.
2) Chainer Framework: Chainer [31], written in Python, can be extended to multiple nodes and GPU platformws through the CuPy and MPI4Python libraries [32][33].
3) DyNet Framework: DyNet [34] was written in C++. The framework can readily define dynamic computation graphs, so DyNet can help improve development speed. Currently, DyNet only supports single nodes and not multiple node platforms.

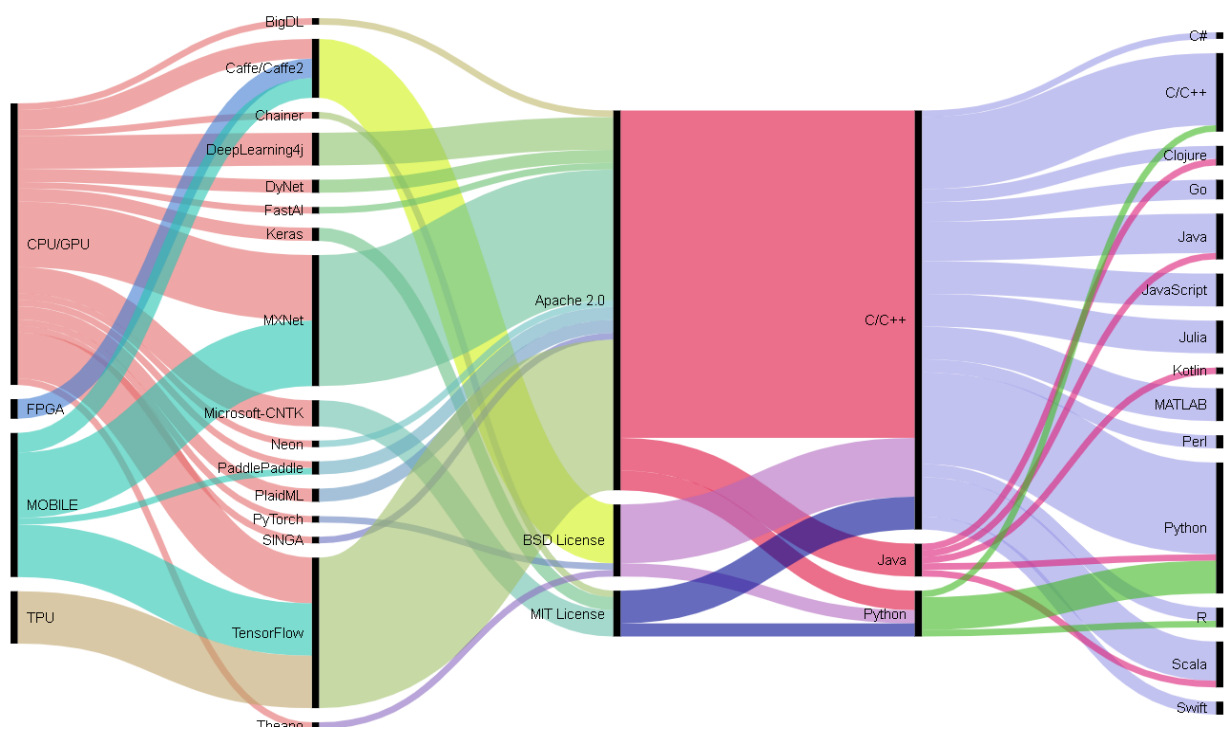

Fig. 4. Popular Deep learning Frameworks. From right column to left one is hardware, frameworks, license types, core codes, and API codes

4) *MXNet:* the Apache MXNet [35][36] *is a well known deep learning framework. This framework was built in C++, and MXNet supports NVIDIA GPUs through the NVIDIA CuDNN library. In* [37], *the GLUNO is a development interface for MXNet.*

5) *Microsoft CNTK: The Microsoft Cognitive Toolkit (Microsoft CNTK)* [38][39], *funded by Microsoft and written in C++, supports distributed platforms.*

6) *Google TensorFlow: In 2011, Google released DistBelief* [40], *but the framework was not an open source project. In 2016, the project was merged with TensorFlow* [41][42], *an open source deep learning framework.*

7) *Keras* [43][44] *is a Python library for TensorFlow, Theano, and Microsoft CNTK. Keras has a reasonable development interface that can help developers to quickly develop demo systems and reduce development costs and risks.*

8) *Neon and PlaidML are partially supported by Intel: Neon* [45], *supported by Nervana Systems and Intel, may improve performance for deep learning on diverse platforms. PLaidML*[46] *was released by Vertex.AI in 2017; Intel will soon fund PlaidML.*

9) *PyTorch Framework: PyTorch* [47][48], *written in Python, can be integrated with Jupyter Notebook. FastAI* [49] *is another development interface for PyTorch.*

10) *Theano Framework: The core language of Theano* [50][51] *is Python with a BSD license. Lasagne* [52][53] *is an additional development library for Theano.*

B. Developing Frameworks

In addition, some deep learning frameworks are less frequently mentioned by academic papers because of their limited functions. For example,

1. Apache SINGA [54] was developed in C++. The framework is supported by the Apache group [44] [45].
2. BigDL [46][47], built with Scale codes, is a deep learning framework that can run on Apache Spark and Apache Hadoops.
3. In [59], the authors mentioned DeepLearning4J (DL4J), which can be accelerated by cuDNN.
4. The PaddlePaddle deep learning framework was developed by Baidu using Python [60].

C. Inactive Frameworks

We mention two of these. (1) Torch [61], was written in Lua. It is inactive. (2) Purine [53][54] is open source and not updated since 2014.

D. Qualitative Benchmarking Metrics for Deep Learning Frameworks

Benchmarking metrics for frameworks for deep learning include six qualitative metrics described next.

1) *License Type: Open source software licenses impose a variety of restrictions. In* [64], *degree of openness is used as a metric for ranking open source licenses. Apache license 2.0 has relatively few restrictions. The MIT license requires the most limitations. BSD is in the middle. So, in comparing degree of openness, Apache 2.0 > BSD > MIT.*

2) *Interface Codes (also called the API): The more functionality the API offers, the better it tends to support development. A good API can increase development productivity, reduce development cost and enhance functionality of the framework.*

3) *Compatible Hardware: Computing hardware devices including CPUs and GPUs constitute the underlying support for deep learning frameworks. The more different hardware devices a deep learning framework can run on, the better it is on this dimension.*

4) *Reliability: No single point of failure (NSPOF) is a risk minimizing design strategy. This approach ensures that one fault in a framework will not break an entire system. For avoiding single points of failure, a mature framework might run on multi-server platforms rather than a single node.*

5) *Tested Deep Learning Networks: Evaluating software could discover potential problems, measure performance metrics, and highlight strengths and weaknesses. If a framework can be officially verified by a variety of deep learning networks, then the framework is correspondingly more suitable as a mainstream production framework.*

6) *Tested Datasets: Image datasets, voice datasets, and text datasets are among those used for training and testing deep learning networks. If a framework was verifed by diverse datasets, we are able to know its performance, strengths, and weaknesses.*

Consistent with these six metrics, there are 16 mainstream deep learning frameworks as shown in Figure 4 and Table II (shown after the references).

IV. A MACHINE LEARNING BENCHMARK ORGANIZATION

MLPerf is a machine learning benchmark organization that offers useful benchmarks that evaluate training and inference on deep learning hardware devices. MLPerf and its members are associated with advanced chip hardware companies and leading research universities. Hardware companies include Google, Nvidia, and Intel. Research universities include Stanford University, Harvard University, and University of Texas at Austin.

MLPerf members share their benchmarking results. Benchmark results, source codes, deep learning algorithms (also called deep learning models), and configuration files are submitted to a website on github.com. Currently MLPerf members already have submitted the MLPerf Training Results v0.5 and MLPerf Training Results v0.6, and the deep learning reference results v0.5 will be released soon.

MLPerf benchmarks involve benchmark metrics, datasets, deep learning algorithms, and deep learning frameworks. MLPerf members execute deep learning algorithms on hardware devices, then record execution time, deep learning algorithms, deep learning frameworks, and tested open datasets. Time is a critical metric for measuring MLPerf training or inference benchmarks [65]. Short run time is associated with high performance of deep learning devices. Benchmark datasets consist of image datasets, translation datasets, and recommendation datasets. ImageNet and COCO [66] are among the image datasets. WMT English-German [67] and MovieLens-20M [68] are translation and recommendation datasets, respectively. MLPerf benchmark frameworks are TensorFlow, PyTorch, MXNet, Intel Caffe, and Sinian. MLPerf deep learning algorithms benchmarked [69] include ResNet50-v1.5, MobileNet-v1, SSD-MobileNet, and SSD-ResNet34.

## V. CONCLUSIONS

Deep learning has increased in popularity dramatically in recent years. This technology can be used in image classification, speech recognition, and language translation. In addition, deep learning technology is continually developing. Many innovative chipsets, useful frameworks, creative models, and big data sets are emerging, resulting in extending the markets and uses for deep learning.

While deep learning technology is expanding, it is useful to understand the dimensions and methods for measuring deep learning hardware and software. Benchmarking principles include representativeness, relevance, equity, repeatability, affordable cost, scalability, and transparency. Major deep learning hardware platform types include CPUs, GPUs, FPGAs, and ASICs. We discussed machine learning platforms, and mentioned approaches that enhance performance of these platforms. In addition, we listed 11 qualitative benchmarking features for comparing deep learning hardware.

AI algorithms often benefit from many-core hardware and high bandwidth memory, in comparison to many non-AI algorithms that are often encountered in practice [70]. Thus it is not just the computational power of hardware as a one-dimensional concept that makes it more (or less) suited to AI applications, but also the type of computations the hardware excels in. A hardware platform can have more or less computational power depending on the type of computation on which it is measured. GPUs (graphics processing units) often do comparatively well on the kind of parallelism often beneficial to AI algorithms, and thus tend to be well suited to AI applications. FPGAs, being configurable, can be configured to perform well on AI algorithms although currently they lack the rich software layer needed to be as useful for this as they could become. ASICs are similar to FPGAs in this regard, since in principle a specially configured FPGA is a kind of ASIC.

Software frameworks for deep learning are diverse. We compared 16 mainstream frameworks through license types, compliant hardware devices, and tested deep learning algorithms. We split popular deep learning frameworks into three categories: mature deep learning frameworks, developing frameworks, and retired frameworks.

Deep learning benchmarks can help link industry and academia. MLPerf is a new and preeminent deep learning benchmark organization. The organization offers benchmarking metrics, dataset evaluation, test codes, and result sharing.

## VI. FUTURE WORK

Deep learning technology including supporting hardware devices and software frameworks is increasing in importance, so scientists and engineers are developing new hardware and creative frameworks. We are planning a website named Benchmarking Performance Suite (http://www.animpala.com/research.html) for collecting and updating results of benchmarking hardware and frameworks. Users will be able to access the website for sharing deep learning knowledge.

ACKNOWLEDGMENT

We are grateful to Google for partial support of this project in 2019.

TABLE II.    COMPARING POPULAR DEEP LEARNING FRAMEWORKS

| # | Frameworks [a] | License Type [b] | Core Codes | API Codes | Hardware Devices | Reliability | Tested Networks | Related Datasets |
|---|---|---|---|---|---|---|---|---|
| 1 | BigDL | Apache 2.0 | C/C++ | Scala | CPU/GPU | Multi-Server | VGG,Inception,ResNet,GoogleNet | ImageNet, CIFAR-10 |
| 2 | Caffe/Caffe2 | BSD License | C/C++ | Python, C++ MATLAB | CPU/GPU/FPGA/Mobile | Multi-Server | LeNet, RNN | CIFAR-10,MNIST, ImageNet |
| 3 | Chainer | MIT License | C/C++ | Python | CPU/GPU | Multi-Server | RNN | CIFAR-10, ImageNet |
| 4 | DeepLearning4j | Apache 2.0 | Java | Java, Scala, Clojure, Python, Kotlin | CPU/GPU | Multi-Server | AlexNet,LeNet,Inception, ResNet, RNN, LSTM, VGG,Xception, | ImageNet |
| 5 | DyNet | Apache 2.0 | C/C++ | C++, Python | CPU/GPU | Single Node | RNN, LSTM | ImageNet |
| 6 | FastAI | Apache 2.0 | Python | Python | CPU/GPU | Multi-Server | ResNet | CIFAR-10, ImageNet |
| 7 | Keras | MIT License | Python | Python, R | CPU/GPU | Multi-Server | CNN, RNN | CIFAR-10,MNIST |
| 8 | Microsoft CNTK | MIT License | C/C++ | C++, C#, Python, Java | CPU/GPU | Multi-Server | CNN, RNN,LSTM | CIFAR-10, MNIST,ImageNet,P-VOC |
| 9 | MXNet | Apache 2.0 | C/C++ | C++, Python, Clojure, Julia, Perl, R, Scala, Java,JavaScript,Matlab | CPU/GPU/Mobile | Multi-Server | CNN, RNN,Inception | CIFAR-10, MNIST,ImageNet,P-VOC |
| 10 | Neon | Apache 2.0 | Python | Python | CPU/GPU | Multi-Server | AlexNet, ResNet, LSTM | CIFAR-10, mnist,ImageNet |
| 11 | PaddlePaddle | Apache 2.0 | C/C++ | Python | CPU/GPU/Mobile | Multi-Server | AlexNet,GoogleNet,LSTM | CIFAR-10, ImageNet |
| 12 | PlaidML | Apache 2.0 | C/C++ | Python, C++ | CPU/GPU | Multi-Server | Inception, ResNet, VGG, Xception, MobileNet, DenseNet, ShuffleNet, LSTM | CIFAR-10, ImageNet |
| 13 | PyTorch | BSD License | Python | Python | CPU/GPU | Multi-Server | AlexNet,Inception, ResNet, VGG, DenseNet, SqueezeNet | CIFAR-10, ImageNet |
| 14 | SINGA | Apache 2.0 | C/C++ | Python | CPU/GPU | Multi-Server | RNN, AlexNet,DenseNet, GoogleNet, Inception, ResidualNet,VGG | MNIST, ImageNet |
| 15 | TensorFlow | Apache 2.0 | C/C++ | Python, C++, Java, Go, JavaScript, Scala, Julia, Swift | CPU/GPU/TPU/Mobile | Multi-Server | AlexNet,Inception, ResNet, VGG, LeNet, MobileNet | CIFAR-10, mnist,ImageNet |
| 16 | Theano | BSD License | Python | Python (Keras) | CPU/GPU | Multi-Server | AlexNet, VGG, GoogleNet | CIFAR-10, ImageNet |

[a.] alphabetical order

[b.] In License Type column, Apache 2.0 means the Apache 2.0 license